\documentclass[a4paper]{article}

\usepackage{INTERSPEECH2020}

\usepackage[utf8]{inputenc} 
\usepackage[T1]{fontenc}    
\usepackage{hyperref}       
\usepackage{url}            
\usepackage{booktabs}       
\usepackage{amsfonts}       
\usepackage{nicefrac}       
\usepackage{microtype}      
\usepackage{lipsum}
\usepackage{color}

\usepackage{hyperref}
\usepackage{url}

\usepackage{bm}
\usepackage{multirow,booktabs}
\usepackage{amsmath}

\title{Speech-to-Singing Conversion based on Boundary Equilibrium GAN}
\name{Da-Yi Wu$^{1,2}$ and Yi-Hsuan Yang$^2$}
\address{
  $^1$Department of CSIE, National Taiwan University\\
  $^2$Research Center for IT Innovation, Academia Sincia}
\email{r07922119@ntu.edu.tw, yang@citi.sinica.edu.tw}

\begin{document}

\maketitle
\begin{abstract}
This paper investigates the use of generative adversarial network (GAN)-based models for converting a speech signal into a singing one, without reference to the phoneme sequence underlying the speech. This is achieved by viewing  speech-to-singing conversion as a style transfer problem.  Specifically, given a speech input, and 
the F0 contour of the target singing output, the proposed model 
generates the spectrogram of a singing signal with a progressive-growing encoder/decoder architecture. Moreover, the model uses a boundary equilibrium GAN loss term such that it can learn from both paired and unpaired data. The spectrogram is finally converted into wave with a separate GAN-based vocoder. Our quantitative and qualitative analysis show that the proposed model generates singing voices with much higher naturalness than an existing non adversarially-trained baseline. 
\end{abstract}

\section{Introduction}

The goal of singing voice synthesis is to create natural-sounding singing voices with some given conditions, such as the lyrics,  pitch labels, or reference audio \cite{saino2006hmm,valle2019mellotron,nakamura19arxiv,lee19interspeech,gu2020bytesing}. 
The reference audio can be a singing passage of a person, and the task is to convert the timbre of the singing passage into the timbre of someone else \cite{eric_sing}. The reference audio can also be a passage of speech voice by someone, and the task is to convert it into a singing passage with the same timbre identity and linguistic content, without reference to the underlying phoneme sequence  \cite{saitou2007speech,vijayan19spm,parekh2020speech}. We are interested in the latter task, i.e.,  \emph{speech-to-singing} (STS) conversion, in this paper, for its 
interesting applications in entertainment, karaoke, music production, and others.


Even through there are many properties shared by speech and singing signals, they cannot be easily converted to one another. Rhythm and phoneme representations in speech and singing are fairly different, and one singing sound can correspond to spoken passages with different speeds, tones, and even pronunciations \cite{vijayan19spm}. Moreover, melody information, less important for speech, is critical and indispensable for the expression of singing. Hence, in addition to preserving the linguistic content and timbre identity, an STS model would also need to generate singing that follows a pre-given or automatically-generated melody contour.

In the literature, there have been two main approaches to STS: model-based and template-based ones. For \textbf{model-based STS}, the work presented by Saitou \emph{et al.} \cite{saitou2007speech} is a representative one. They decompose the STS process into three main parts and process the signal using different control models. This involves lengthening the speech by a duration control model, generating the F0 contour with an F0 model, and modifying the timbre of the voice (so that it is singing-like) by a spectral model. Yet, the synthesis quality much depends on how accurate the phonemes are segmented and associated with the musical notes. 

For \textbf{template-based STS},  proposed for the first time in \cite{cen2012template}, assumes that a high quality-singing vocal, a.k.a., the ``template singing,'' is available as another audio reference. The inputs therefore comprise the 
speech and the template singing, which are to be firstly aligned with one another. The template singing is further used to extract the reference prosody which includes singing F0, aperiodicity index, singing formants, etc. This information is then used to estimate the parameters of singing synthesis from the aligned speech. As another example, Gao \emph{et al.} \cite{gao2019speaker} propose a deep learning based model for template-based STS conversion which conditions the network on the i-vector of a speaker while predicting the singing spectral parameters to preserve the speaker identity.

In this paper, we view STS as a \textbf{style transfer} problem, which can be considered as the third approach to STS. Speech and singing are two very different styles, in that 
in singing we care more about  melody and rhythm. 
Among the existing style transfer architectures, 
we adopt the GAN \cite{zhu2017unpaired} architecture for its generalizability, and demonstrated effectiveness in  other musical style transfer tasks \cite{eric_sing,music_translation}.

This work represents a continuation and extension of our prior work presented in  \cite{parekh2020speech}, which is among the first attempts to approach STS with a deep-learning model that does not require any phoneme synchronization information or high-quality singing reference. In other words, only a speech passage and a target melody (F0) contour is needed for this model. We show in \cite{parekh2020speech} that even there is no other additional information, the model can still learn to sing, albeit the quality of synthesis is not good enough. The limited quality of the generated singing can be attributed to many parts of the model, and it is our goal to improve upon this prior art here by introducing adversarial learning and modern style transfer techniques.



Specifically, our contributions are as follows. First, we replace the straightforward convolution architecture of \cite{parekh2020speech} by an architecture that progressively grows the output spectrogram with a hierarchical structure. Such an idea has been increasingly used to generate  audio in  recent work \cite{karras2017progressive, vasquez2019melnet,aa,binkowski2020High}. Second, instead of relying fully on paired data as in \cite{parekh2020speech}, we extend the model so that it can also learn from unpaired data. 
Specifically, an adversarial architecture based on boundary-equilibrium GAN \cite{berthelot2017began} is employed. 
Third, we adopt the `random resampling' idea proposed in the autoVC paper \cite{AutoVC} to better disentangle content and rhythm information. 
Finally, we use a MelGAN-based neural vocoder \cite{melgan} to replace the Griffin-Lim algorithm \cite{griffinlim} adopted in \cite{parekh2020speech} to further improve the sound quality.

For paired data, we use the  the NUS sung and spoken lyrics corpus \cite{nus48e},\footnote{\url{https://smcnus.comp.nus.edu.sg/nus-48e-sung-and-spoken-lyrics-corpus/}} which is the largest public paired dataset to date. For unpaired data, we use the DAMP corpus \cite{damp},\footnote{\url{https://ccrma.stanford.edu/damp/}}  which is comprised of performances of 5,690 unique popular songs. 

We report both quantitative and qualitative evaluation to compare the proposed model with the prior art \cite{parekh2020speech} as well as ablated versions of our model. 
Open source implementation of the proposed model, as well as audio examples of the generated conversions can be found online.\footnote{\url{https://github.com/ericwudayi/speech2singing}}$^,$\footnote{\url{https://ericwudayi.github.io/Speech2Singing-DEMO}}

\begin{figure}[t]
  \centering
  \centerline{\includegraphics[width=8.0cm]{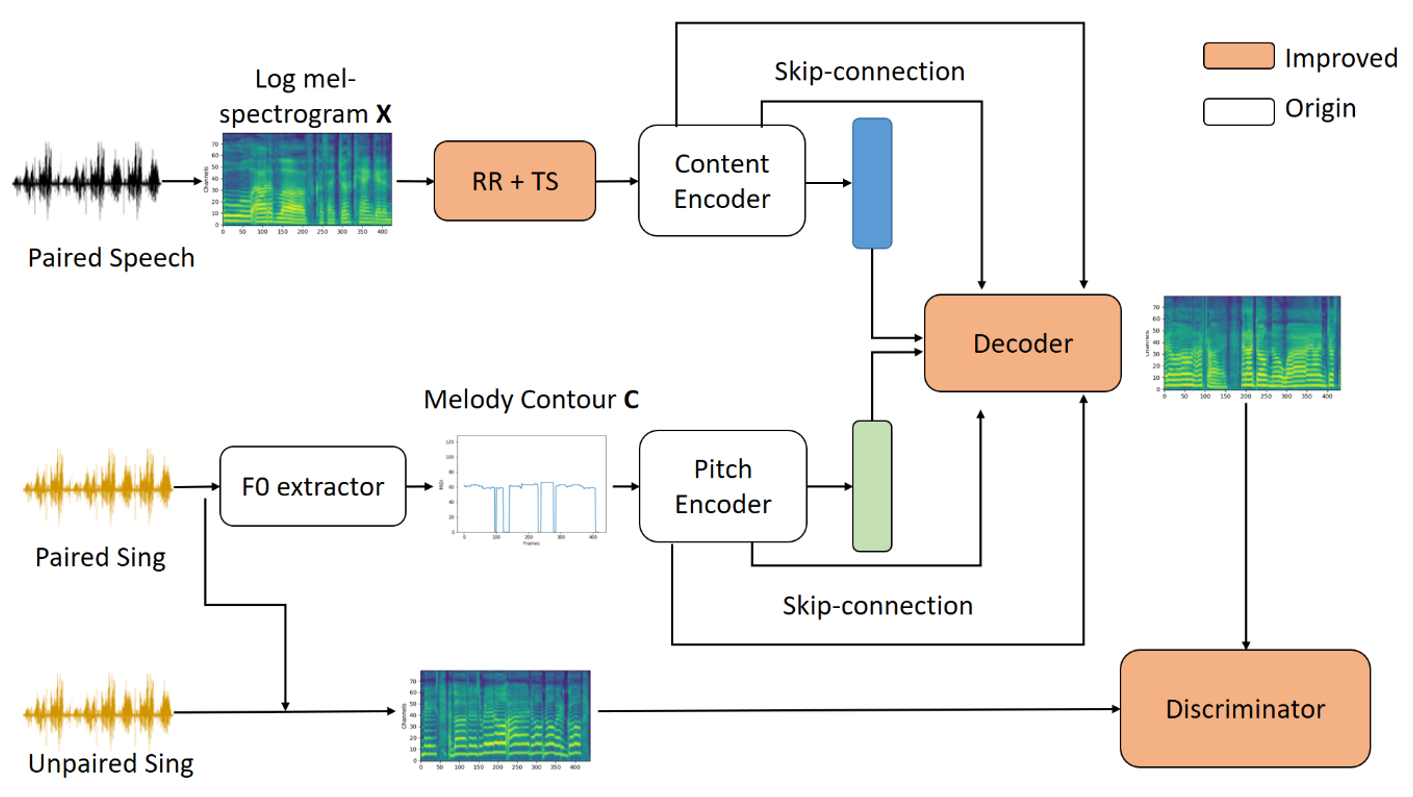}}
\caption{Diagram of the proposed model architecture, which is extended from the model we previously presented in  \cite{parekh2020speech}. The input audio is represented by a log mel-spectrogram, and the target melody contour is represented by the one-hot format. The model learns to perform STS conversion by joint supervised and unsupervised learning. We highlight the components that are different from the prior art \cite{parekh2020speech} in orange. `RR' stands for random resampling \cite{AutoVC}, while `TS' stands for time stretching.}
\label{fig:fig1}
\end{figure}

\section{Methods}
We perform spectra-to-spectra conversion in an encoder-decoder framework as depicted in Fig. \ref{fig:fig1}. The input speech is transformed into a log mel-spectrogram, and the F0 contour is extracted by a vocal melody extractor from a different source such as humming or reference singing. We time-stretch the mel-spectrogram of speech to the same length as its target F0 contour, and apply two encoders to encode speech and pitch information separately. Then, the decoder takes the concatenation of these two encodings and generates the singing output, with skip-connections from the encoder. Finally, we use MelGAN vocoder to reconstruct the waveform from the log mel-spectrogram. Additionally, we add BEGAN discriminator trained on both model output and ground truth to improve the model generalizability. The whole model is trained on both paired and unpaired data. We explain below each of the components in more detail.


\begin{figure}[t]
  \centering
  \centerline{\includegraphics[width=8.0cm]{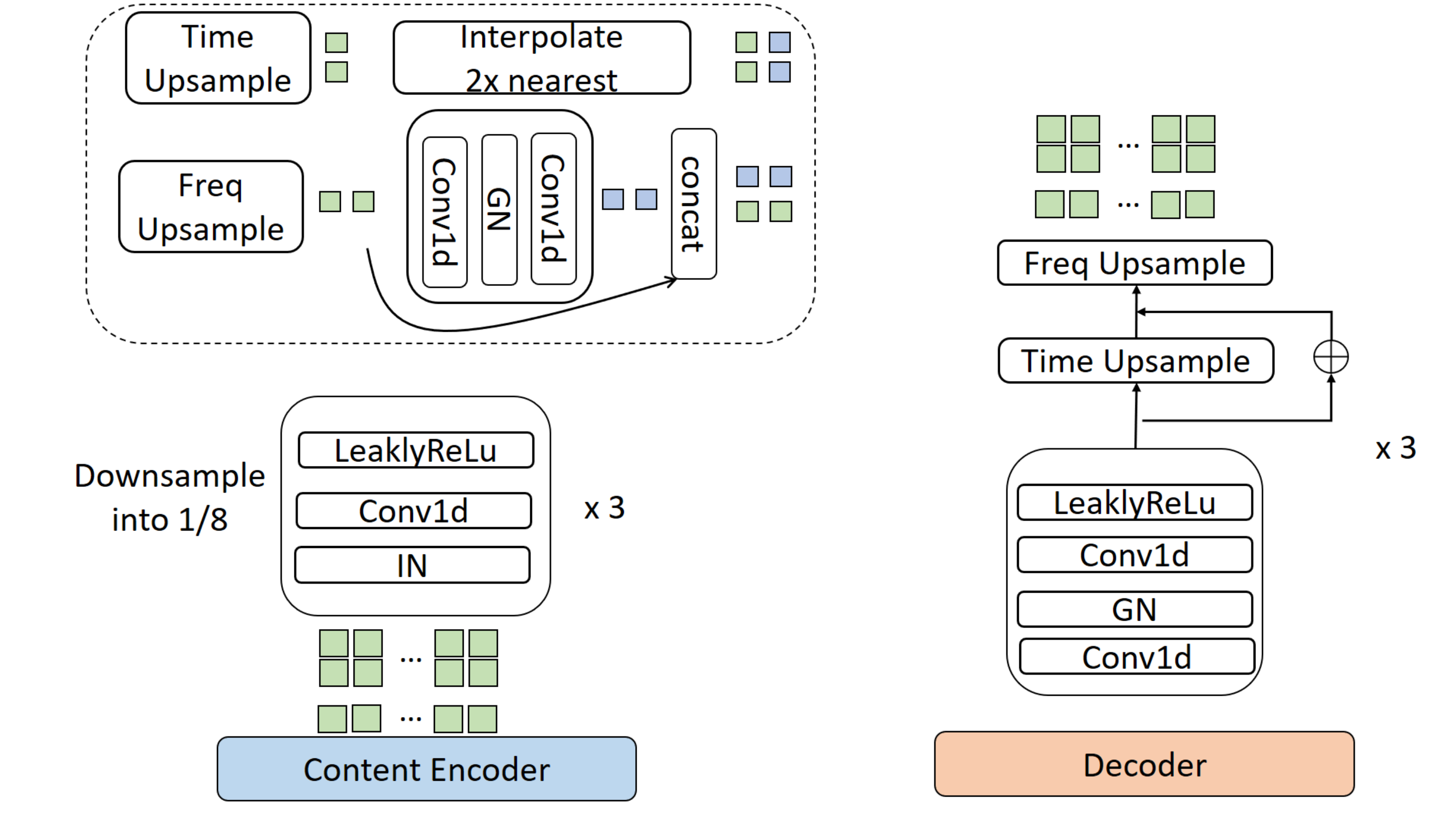}}
\caption{Details of the employed encoder and decoder. The encoder consists of convolution banks, and the decoder upsamples the time-frequency representation by interpolation and concatenation. `IN' and `GN' represent instance normalization and group normalization, respectively. Details of time and frequency upsample blocks are pictured on the top-left side.}
\label{fig:fig2}
\end{figure}

\subsection{Input processing}
Given an input time-domain speech signal and a target F0 contour, the pre-processing consists of the following steps.

    \textbf{Log-magnitude representation}: We compute the magnitude of mel-spectrogram for the speech signal and apply element-wise logarithm transformation, 
    yielding a matrix $\mathbf{X} \in\mathbb{R}^{F \times T}$, with $F$ frequency bins and $T$ time frames.
    
  \textbf{Random resampling (RR)}: A speech passage can be ``sung'' with various speed and rhythmic characteristics.  Following autoVC \cite{AutoVC}, we random resample the input speech to change the rhythm of the input speech, so as to  better disentangle content and rhythm. This is done by firstly dividing the mel-spectrogram of speech into segments of random lengths, and then randomly stretching or squeezing each segment along the time axis. In our implementation, we divide the speech into segments of random length of 16--32 frames, and temporally stretch  each segment by a factor of 0.5--2.
  
  \textbf{Singing melody contour} is extracted from a singing audio using  CREPE \cite{kim2018crepe}, a state-of-the-art, open source monophonic pitch tracker. Moreover, we convert the continuous-valued extracted F0 to one of the 128 MIDI notes by referencing to the Hz-to-MIDI function in the librosa package \cite{mcfee2015librosa}, and convert it into the one-hot format. As a result, the melody contour $\boldsymbol{C}$ is represented as a sequence of one-hot vector indicating the target MIDI note per frame, namely $\boldsymbol{C} \in N^{128 \times T'}$. Following \cite{parekh2020speech}, in the case of training with paired data, we extract the melody contour from the singing counterpart of the input speech.
  
  \textbf{Time stretching (TS)}. The speech mel-spectrogram and the target F0 contour (which is extracted from a singing signal) are usually much different in length. Therefore, it would be better if our model does not assume that the input and output are of the same length. 
  However, existing models  addressing the variable-length task, such as the sequence-to-sequence model Tacotron \cite{tacotron2}, often deal with only one of the following two cases: 1) both the input and output are discrete, or 2) the input is  discrete while the output is continuous. Spectra-to-spectra learning tasks like STS have to deal with the case where both the input and output are continuous, something that has not been widely studied yet. Therefore, we use a simpler \textbf{fixed-length setting} (and left the variable-length version as a future work). Accordingly, the input mel-spectorgram is linearly interpolated  to $\mathbf{X} \in\mathbb{R}^{F \times T'}$, i.e., to be of the same length as the F0 contour.

  
\subsection{Encoder and Decoder}
As shown in Fig. \ref{fig:fig2}, the content encoder uses 3 stacks of layers to obtain the latent code, downsampling both time and frequency by a factor of 8. Each stack is composed of a instance normalization layer \cite{IN} (as a ``style remover''), 
a 1D-convolution layer (with $3 \times 1$ convolution kernels with stride $2$; no pooling), and a LeakyReLu activation layer. On the other hand, the pitch encoder (not shown in Fig. \ref{fig:fig2}) applies embedding layer to project the one-hot vectors $\boldsymbol{C}$ to an embedding space first, and then passes it through similar 1D-convolution based layers. 

As for the decoder, we use a progressive-growing architecture, which has been recently shown to be an effective ``spectrogram generator'' recently \cite{vasquez2019melnet,aa}. Specifically, it uses 3 stacks of layers, each stack comprising two 1D $3 \times 1$ kernel convolutions with stride $1$, a Group-Norm layer \cite{wu2018group}, and two upsampling modules. Following UNAGAN \cite{aa}, nearest neighbor interpolation is applied instead of 1D transposed-convolution. And, Upsampling in frequency domain is performed by concatenation instead of a convolution layer, in the similar light of  \cite{vasquez2019melnet,aa}. 

\subsection{BEGAN}
BEGAN is an energy-based GAN architecture \cite{berthelot2017began}. While
the original GAN \cite{goodfellow14} matches the distributions between the real and
generated samples directly, an energy-based GAN matches the distribution of loss using an auto-encoder architecture. Moreover, BEGAN relaxes the equilibrium of the auto-encoder loss using a hyper-parameter $\gamma \in [0,1]$ (a.k.a, diversity ratio),  defined as \begin{equation}
    \gamma = \mathbb{E}[L(G(\mathbf{x}))] / \mathbb{E}[L(\mathbf{y})] \,,
\end{equation}
where $G(\mathbf{x})$ is a generated sample, $\mathbf{y}$ is a real sample, and $L(\cdot)$ is the reconstruction loss function of the auto-encoder. The $\gamma$ term balances the diversity and quality. Lower values of $\gamma$ make the discriminator focuses more on real samples, generating samples with better quality but less diversity. In contrast, higher values of $\gamma$ improve the diversity but lower the quality.

We use BEGAN as our unpaired data trainer for its stable training process observed in our pilot study. It has also been shown in \cite{wu18singing} and \cite{liu2019score} that BEGAN performs better than some other commonly-used GAN models for generating audio.

\subsection{Training}
Our model is trained with by minimizing the BEGAN loss and L1 loss jointly. Formally, given an input log mel-spectrogram $\mathbf{X}$, a melody contour $\mathbf{C}$, and a target singing log mel-spectrogram $\mathbf{Y}$, the losses of the generator and discriminator are defined as:
\begin{equation}
\left\{
\begin{aligned}
L_D &= L(\mathbf{Y}) - k_tL(G(\mathbf{X}, \mathbf{C})) \,,
\\
 L_G &=  L(G(\mathbf{X}, \mathbf{C})) + \beta (| \mathbf{Y} - G(\mathbf{X}, \mathbf{C}) |_1) \,, \label{eq:2}
\end{aligned}
\right. 
\end{equation}
where $L_D$ only depends on BEGAN loss, and the variable $k_t \in [0, 1]$ controls how much emphasis is put on $L(G(x))$. The update of $k_t$ is controlled by the diversity ratio $\gamma$ according to
\begin{equation}
k_{t+1} = k_t + \lambda(\gamma L(\mathbf{Y}) - L(G(\mathbf{X}, \mathbf{C}))) \,.
\end{equation}
In Eqn.~(\ref{eq:2}), we also use the pixel-wise $L1$ spectrogram loss to $L_G$ to achieve supervised learning in paired data, scaling with a $\beta$ factor to avoid overfitting. As for the case of training with unpaired data, the generator is trained with $L(G(\mathbf{x}))$ only.  

\subsection{Vocoder}

Neural vocoders such as the WaveNet vocoder \cite{oord2016wavenet} have been shown to outperform the traditional Griffin-Lim algorithm \cite{griffinlim} for converting time-frequency representations to  waveforms.
Among the neural vocoders that have been proposed, we adopt MelGAN \cite{melgan} for its remarkable efficiency and generalizability.

\section{Experimental Validation}
\label{spsi-ev}
\subsection{Setup}

\textbf{Training and test sample generation}. 
We employ the NUS database \cite{nus48e} as our paired training dataset, which contains 115 mins of singing data and the corresponding 54 mins of speech data. 48 recordings of 20 unique English songs are sung and read out by 12 subjects, and each sung-read paired audio can be time-aligned by their phone-duration annotations. 
The phone annotations is in accordance to the CMU phoneme dictionary (39 phonemes)\footnote{\url{http://www.speech.cs.cmu.edu/cgi-bin/cmudict}} with two extra phones denoting the silence and inhalation between words, and the duration annotations specify the start and end time of each phone. 
We remove the silence from speech by using the phoneme-duration annotation.  All the speech input are constrained to have three or more words. 

Out of the 20 unique songs in the dataset, we keep one song (with two recordings) as our test set. The test singer is only present in one recording in our training set, and the test song is not seen at all in the training data. 

Due to the small size of the NUS dataset, we perform data augmentation using unpaired data with the DAMP dataset \cite{damp}, a multi-singer, singing-only dataset comprising of performances of 5,690 unique songs by 13,154 users of a karaoke App (i.e. a song can be sung by different people).
We exclude the segments of song that contains silences of longer than one second.\footnote{Our originally plan is to also use LJSpeech \cite{ljspeech17}, a single-speaker speech-only dataset with 24 hours of reading audio for data augmentation. We later realize that this is not easy, as in our STS setting, we need to find a way to create the target melody contour for each input speech. 
The melody contour should correlate well to the number of words from the spoken lines. And, it is not natural to convert the speech with arbitrary melody contour. We attempt to give the pitch contour by randomly picking a pitch contour from DAMP, but the result shows that it actually hurts the audio quality.
We therefore finally decide not to use LJSpeech.}

\textbf{Implementation details}. All the audio processing procedures are performed using librosa \cite{mcfee2015librosa}. The time domain signals 
are resampled from 44k to 22k Hz. We compute STFT with 1,024-pt FFT size, 12.5 milliseconds hop size, and transform it into 80-bin mel-scale.  We set BEGAN parameters $\lambda = 0.01$ and initialize $k_0 = 0$, $\gamma = 0$.  We set supervised learning factor $\beta = 0.5$ and use Adam as our optimizer with initial learning rate 0.001 and exponential decrease factor of 0.99. The
neural network is implemented using pytorch. We train the networks for 20k steps with 32 batch size. The training process takes about 5 hours to complete on an NVIDIA RTX 1080-Ti GPU. 

Our multi-singer MelGAN vocoder is trained on the union of the NUS dataset \cite{nus48e} and two other sources of unaccompanied singing data, the DSD \cite{SiSEC16} and MUSDB18 datasets \cite{musdb18}.\footnote{These two datasets provide unaccompanied singing so there is no need to perform singing/accompaniment source separation \cite{liu19ijcai}.} The whole training process take 1000k steps, for about 2 weeks on the same 1080-Ti GPU. 

\textbf{Phoneme synchronization (PhSync)}. 
To quantify how much the burden of modelling phoneme duration affects STS performance, as an \emph{oracle} method we follow the settings in \cite{parekh2020speech} and stretch each phone in the input speech to be the same length as the target singing. The duration for each phoneme is obtained from the phone-level annotations for speech and singing. 

\subsection{Models evaluated}
We evaluate several ablated versions of our model to study the effect of the proposed changes.
\begin{itemize}
    \item \emph{Baseline}: Uses the model architecture proposed in the prior art \cite{parekh2020speech}, which is trained through multi-task learning with phoneme prediction, using only the paired data. It uses log spectrograms as its input, not mel-spectrograms. 
    \item \emph{Decoder} \textbf{(D)}: Uses the modified version of the decoder, and trained with the MSE loss. We use the spectrograms here as the input for fair comparison with the Baseline.
    \item \emph{Decoder + Adversarial} \textbf{(D+A)}: Uses the modified version of the decoder, and trained with MSE and adversarial losses jointly. Also use the spectrogram features. In other words, the unpaired data is used.
    \item \emph{Decoder + Neural Vocoder} \textbf{(D+V)}: Uses our modified version of the decoder, and trained instead on the log mel-spectrogram features. The output is converted to waveform by MelGAN~\cite{melgan} instead of Griffin-Lim. 
    \item \emph{Complete model} \textbf{(D+V+A)}: The proposed model trained with adversarial loss and MSE loss jointly using both paired and unpaired data, and the output is also converted into waveform by MelGAN.
    \item \emph{Complete model + PhSync}:  The proposed model trained in the oracle phoneme synchronization setting.
\end{itemize}

\begin{table}
\centering
\begin{tabular}{l c c c c} 
\toprule
Model  &  LSD $\downarrow$ & LSD (mel) $\downarrow$ & RCA $\uparrow$\\ \toprule
Baseline \cite{parekh2020speech} & 9.97  & --- & 0.760 \\
\textbf{D} & 9.36  & --- & 0.801\\ %
\textbf{D+A} & \textbf{9.21}  & --- & \textbf{0.820}\\ 
\midrule
\textbf{D+V} & --- & 1.15 & 0.811\\ 
\textbf{D+V+A} & --- & \textbf{1.13} &  \textbf{0.832}\\ 
\midrule
\textbf{D+V+A} + PhSync & --- & 1.07  & 0.816\\ 
\bottomrule
\end{tabular}
\caption[Results]{
Results on objective metrics for different proposed STS models. 
Log-spectral distance (LSD; in db) is the lower the better, while raw chroma accuracy (RCA) the inverse. 
The first three models take spectrogram as input and use Griffin-Lim for inverse STFT, while the others use mel-spectrorams and MelGAN-based vocoder. Therefore, `LSD' and `LSD (mel)' are calculated over magnitude spectrograms and mel spectrograms, respectively. 
We highlight the best result obtained, excluding the last model as it uses oracle phone-annotation (PhSync) information. 
}
\label{results}
\end{table}

\subsection{Objective Evaluation}

Following \cite{parekh2020speech}, we use log-spectral distance (LSD) and F0 raw chroma accuracy (RCA) \cite{mir_eval} as the objective metrics. 
LSD evaluates the dissimilarity between two spectra, while RCA evaluates melody correctness. 
LSD is computed by averaging the Euclidean distance between true and predicted log spectrogram or log mel-spectrogram frames over time, for frequencies between 100--3500 kHz. 
RCA is computed between a target waveform and the model output, and we set the maximum tolerance deviation between target and output as 50 cents \cite{mir_eval}. 
The models are evaluated by selecting random 50 test samples with speech duration of at least 2 seconds and then averaging the scores over all the samples.

The result is shown in Table \ref{results}.
Several observations can be made.
First, \textbf{D} performs noticeably better than Baseline on both LSD and RCA. This indicates that our decoder architecture combined with the progressive-growing architecture works better for STS. Second, the models with adversarial learning consistently outperform the non-adversarial counterparts (i.e., \textbf{D+A} outperforms \textbf{D}, while \textbf{D+V+A} outperforms \textbf{D+V}). This suggests that unsupervised learning with the BEGAN architecture improves the result. Finally, in terms of RCA, the best result (0.832) is obtained by the complete model (\textbf{D+V+A}).\footnote{We note that the first three methods take the magnitude spectrograms as inputs, while the other methods deal with the mel-specgrograms. Therefore, the values of  `LSD' and `LSD (mel)' are not comparable.} 

\subsection{Subjective Evaluation}

Instead of doing subjective evaluations on all the models, we conducted preference listening test on the following three selective pairs of  models: `\textbf{D} vs. Baseline,' `\textbf{D+V} vs. \textbf{D},' `\textbf{D+V+A} vs. \textbf{D+V},' to respectively investigate: 1) Is our decoder architecture better than the prior one employed in \cite{parekh2020speech}? 2) Does the neural vocoder with the 80-bin mel-spectrogram performs better than the Griffin-Lim algorithm with the 512-bin spectrogram?  3) Does adversarial learning help generate more natural-sounding singing?  
Each participant was first asked to listen the input speech and then compare the outputs of a given pair of models. The participants were then asked to specify their preference among the two models in terms of naturalness. For all pairs of models, the percentage of votes (including ties) are reported.


\begin{figure}[t]
  \centering
  \centerline{\includegraphics[width=8.0cm]{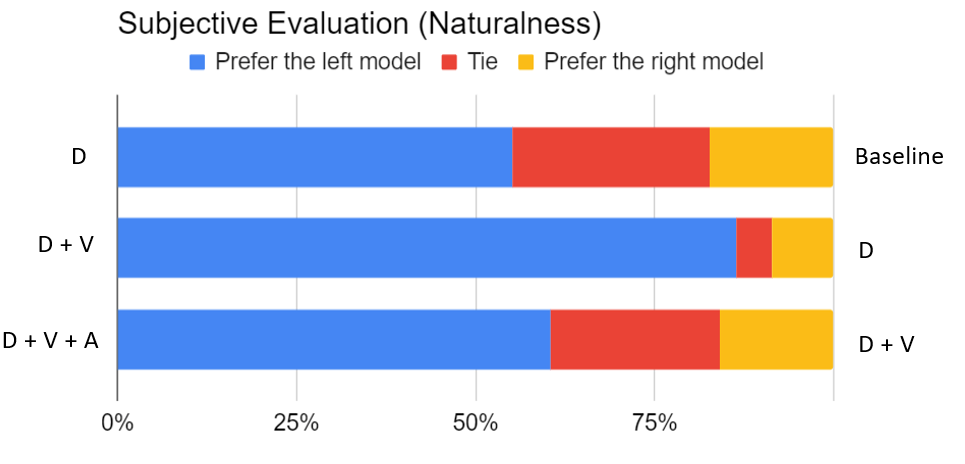}}
\caption{Subjective evaluation results for pairs of systems, from top to bottom:  `D vs. Baseline,' `D+V vs. D,' `D+V+A vs. D+V.' }
\label{fig:downconv}
\end{figure}

The response from 38 participants recruited from the Internet for our listening test is summarized in Fig. \ref{fig:downconv}. Each of our modifications seems to outperform the corresponding ablated version. Key takeaways are: We can do better STS by using a neural vocoder on mel-spectrograms, and the use of adversarial learning and unpaired data is beneficial. 


\subsection{Other Experiments and Future works} 
As the target melody may not be always  available, it would be great if the STS model can generate the melody contour on its own given the speech audio. Hence, in a pilot study we attempt to directly transform speech to singing without pre-specified melody. The cycle-BEGAN architecture is employed to achieve unsupervised learning, and models are trained with adversarial loss and cycle reconstruction loss jointly. However, our preliminary result is not positive; our model does not learn to generate coherent melody across time.
We plan to study in our future work a two-step model that generates the melody contour from speech first, and then generates the singing.
Another idea is to use an auto-regressive models such as the Transformers \cite{vaswani2017attention,huang20remitransformer,dhariwal2020jukebox} 
so that the output is conditioned on previous states.    

\section{Conclusions}
In this paper, we have presented various  modifications that improve an end-to-end speech-to-singing conversion neural net. The model takes only speech and target melody contour as the input. It learns from both paired and unpaired data to perform the conversion. We have also report objective and subjective evaluations to validate the effectiveness of the proposed improvements. In the future, we plan to employ sequence-to-sequence models to tackle the task in the variable-length setting,  preferably without pre-specified singing melody contours.

\bibliographystyle{IEEEtran}

\bibliography{mybib}


\end{document}